# COMPARISON OF CSMA BASED MAC PROTOCOLS OF WIRELESS SENSOR NETWORKS

Himanshu Singh[1] and Bhaskar Biswas[2]

[1]Department of Computer Engineering, IT-BHU, Varanasi, India.
himanshu.singh.cse07@itbhu.ac.in
[2] Department of Computer Engineering, IT-BHU, Varanasi, India.
bhaskar.cse@itbhu.ac.in

## ABSTRACT

*Energy conservation has been an important area of interest in Wireless Sensor networks (WSNs). Medium Access Control (MAC) protocols play an important role in energy conservation. In this paper, we describe CSMA based MAC protocols for WSN and analyze the simulation results of these protocols. We implemented S-MAC, T-MAC, B-MAC, B-MAC+, X-MAC, DMAC and Wise-MAC in TOSSIM, a simulator which unlike other simulators simulates the same code running on real hardware. Previous surveys mainly focused on the classification of MAC protocols according to the techniques being used or problem dealt with and presented a theoretical evaluation of protocols. This paper presents the comparative study of CSMA based protocols for WSNs, showing which MAC protocol is suitable in a particular environment and supports the arguments with the simulation results. The comparative study can be used to find the best suited MAC protocol for wireless sensor networks in different environments.*

## 1. INTRODUCTION

A 'Wireless Sensor Network' (WSN) can be described as a network of sensors which communicate with each other wirelessly. These sensors may be installed in an unattended environment with limited computation and sensing capabilities. Hence, they need to be fault-tolerant and reliable so that maintenance requirement is less. Since sensors are often deployed in remote applications like forest-fire monitoring and structural health monitoring, the battery cannot be replaced frequently due to inaccessibility of sensor nodes. To prolong network lifetime, energy spending should be minimum. Energy conservation can be done using efficient macro-programming of WSNs. In [4], we present a study of methods of energy efficient macro-programming. Another approach is to design energy efficient MAC protocols. Various MAC protocols have been designed to address this issue.

According to a survey [2] on Wireless Sensor Networks, major sources of energy waste at medium access communication are a) collision - which requires re-transmission of collided packets, b) overhearing - where a node receives a message meant for another node, c) control packet overhead - where energy is consumed in exchange of control packets used for control data transmission and d) idle listening - which means that node is listening to idle channel and then over-emitting by sending packets when the destination node is not yet ready.

Among all reasons mentioned above idle listening is a major cause of energy waste. So it is important to design a suitable MAC protocol which can reduce or prevent above energy wastes. There are four techniques to avoid idle listening - static sleep scheduling, dynamic sleep scheduling, preamble sampling, and off-line scheduling. Based on these techniques, many MAC protocols based on CSMA, TDMA, hybrid and cross-layer optimizations were introduced. We classify all the famous MAC protocols for WSNs in Table 1.

Table 1 Classification of Medium Access Control Protocols

| CSMA based(contention based) | TDMA based(Reservation based) |
|---|---|
| SMAC,TMAC,PCSMAC,BMAC,WiseMAC, DMAC,UMAC,XMAC,PMAC,CMAC etc. | ERMAC,TRAMA,EMACS,DEMAC, BMA,SS-TDMA,LMAC etc. |
| Hybrid(CSMA AND TDMA) | CROSS-Layer |
| IEEE 802.15.4,PTDMA,DEE-MAC,$\mu$MAC, SCP-MAC,RMAC,AMAC,SPARE-MAC, YMAC,ZMAC,HMAC etc. | MAC-CROSS,LESOP-MAC |

The MAC protocols behave differently under different network scenarios with respect to energy consumption and throughput. So there is a need of an efficient comparative study of these protocols. In this work, we evaluate CSMA based MAC protocols. The MAC protocols were implemented on TOSSIM [6]. Unlike other simulators (OMNet++ and NS-2), TOSSIM simulates exactly the same code which is going to be executed on real hardware, thus narrowing the gap between simulation and real network deployments. Each MAC protocol is then integrated with PowerTOSSIM-Z [7], a power modelling tool, to measure the energy consumption. In Section 2 to Section 9, we describe the implemented MAC protocols, along with their pros and cons supported by simulation results. Section 10 describes the simulation setup and the results. Section 11 concludes the paper highlighting our future aims and scope.

## 2. S-MAC

Sensor-MAC (SMAC) [5] is a contention-based protocol that regulates sleep periods in a sensor network to conserve energy and improve network lifetime. This protocol represents the baseline of sleep-oriented, energy-efficient WSN MAC protocol designs. Out of four techniques for avoiding idle listening: static sleep scheduling, dynamic sleep scheduling, preamble sampling, and off-line scheduling, SMAC adopts static sleep scheduling for preserving the energy. SMAC divides the time into frames. Every frame is divided into an active and a sleep period as shown in Figure 1. In active period, the transmitter-receiver is switched on and it is switched off during sleep period. The active period is further divided into Time Synchronization period and data transfer period.

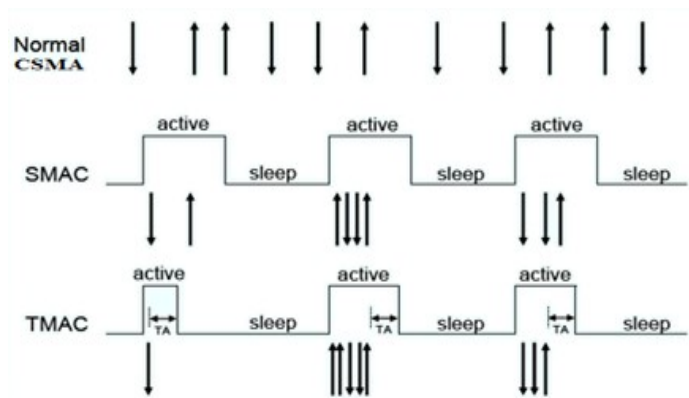

Figure 1: Sleep and wake-up cycles in SMAC and TMAC

Time Synchronization is required so that receiver remains awake when sender sends the message. In Time Synchronization Period, first step in setting the sleep schedule for a node is to listen for a SYNC packet from a neighbour. The SYNC packet contains the sleep schedule and indicates that the sender is going to sleep after 't' seconds. Once the node receives its neighbour's sleeping schedule, it adopts that schedule and re-transmits the schedule for other neighbouring nodes to adopt. If a node does not receive a SYNC packet within a pre-decided timeout period, the node will set and broadcast its own schedule. Border nodes (nodes between two active schedules) may receive two different schedules from different nodes. They may either adopt both or one of the schedules. We implemented SMAC where border nodes follow both the schedules. Due to this mechanism, the network gets divided into multiple virtual clusters, each cluster surrounded by border nodes. Each node within a cluster follows same sleep schedule, whereas border nodes follow schedule of both its neighbouring clusters. Hence, border nodes remain awake for a larger period, thus increasing the energy consumption. However, in every cycle border nodes and virtual clusters keep changing. So virtual clustering does not affect the network lifetime as a whole.

In SMAC, data exchange takes place in data transfer period through RTS-CTS-DATA-ACK handshaking for unicast communication. A node may extend its active duration if data exchange doesn't finish in active period. However, even if data exchange finishes within active period, the node will still remain awake until its sleep time thus wasting energy. When a node sends a RTS (Request to Send) or CTS (Clear to Send), it puts the duration of data transmission in RTS or CTS packets. Neighbouring nodes which overhear these RTS or CTS set an NAV (Network allocation vector) interrupt timer based on the duration and go to sleep since they cannot communicate in this duration due neighbour's disturbance. When NAV interrupt fires, these neighbouring nodes again start following their normal schedule. This handshaking technique not only reduces collision but also saves energy by overhearing avoidance. Figure 5 demonstrates this observation as the energy consumption is low at higher traffic and increases with increase in traffic. This is because neighbour nodes sleep for longer time on overhearing RTS or CTS, frequency of which is higher in high traffic. However, such conservation is not achieved in Broadcast messages, since broadcasting does not use RTS-CTS handshaking.

## 2. T-MAC

Timeout MAC (TMAC) [8] is also a contention-based, MAC layer protocol that is based upon the basic features of SMAC in optimizing power efficiency by sleeping during periodic network inactivity. However, unlike SMAC, TMAC follows dynamic sleep schedule. The TMAC protocol introduces an active timeout mechanism that decreases the idle listening overhead by dynamically adjusting the active period according to network traffic loads. TMAC allows the nodes to sleep after sometime when all network traffic has completed, as explained in Figure 1. The end of traffic is signalled after monitoring an idle channel for an adaptive timeout (TA) period. If no activity occurs for this TA time duration, node switches off its radio and goes to sleep state. The TA period should be large enough to overcome the early sleeping problem (a node goes to sleep state when a neighbour still has packets to be sent). Such procedure makes TMAC more energy savvy than SMAC as evident from Figure 5.

The TA duration depends on contention interval, length of RTS packet and Turnaround time. We find the appropriate TA time for our implementation as shown in Figure 2.

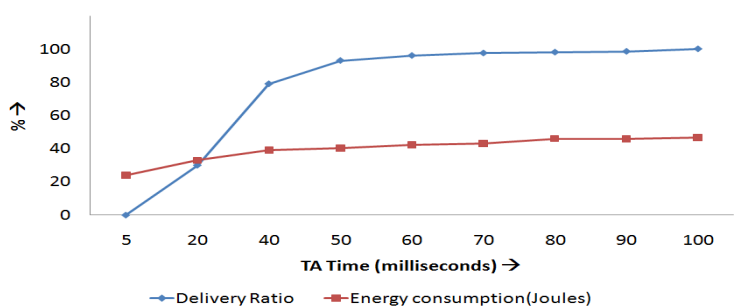

Figure 2. Effect of variation of TA time on delivery ratio and energy consumption

The graph presents the trade-off analysis between energy consumed and delivery ratio, as TA is varied. The non-zero y intercept of graph shows that an appreciable amount of energy is consumed even when there is no data traffic. This energy is consumed during time synchronization period. So, Synchronization is a large overhead in SMAC and TMAC.

TMAC also introduces a FRTS (Future request to send) mechanism and full buffer priority, to avoid early sleeping problem for converging type of data communication. When a node which has a data to send overhears a CTS packet, it broadcasts a FRTS. The duration of the data is stored in FRTS packets. The recipient of FRTS sets its NAV and goes to sleep. After the communication, node again wakes up to receive the data the sender of FRTS. In full buffer priority, when the sending buffer of a node is full and it receives a RTS from some node, then instead of replying with CTS, node transmits its own RTS thus taking the priority. When it has completed the data sending, only then it replies with CTS to the original RTS request it received. So, this mechanism also introduces the flow control in the data flow. It should be noted that Full buffer priority should be employed only in converging type of mechanism and not in ad-hoc type of communication.

## 4. D-MAC

D-MAC [8] is a protocol which aims at real-time delivery of data, still being energy efficient. It adopts a staggered wake-up pattern to forward the data packets to the base-station as shown in Figure 3. Nodes are considered to be present at different levels of data-gathering tree.

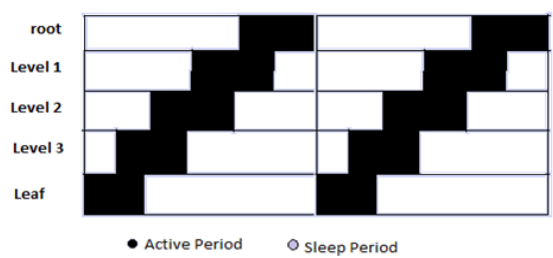

Figure 3. Active and Sleep Period of DMAC

All the nodes at one level would wakeup simultaneously to receive the data. This receiving period, µ, is followed by the transmitting period (µ) in which they forward data to higher level. The nodes at next level wake-up just after the receiving period of the lower level. So, active period is a staggered wake-up pattern, where active period of one level partially overlaps with that of lower level as shown in Figure 3. Due to such staggered wake-up pattern, a data packet reaches from root to leaves in one cycle only, thus minimizing latency. D-MAC adopts data prediction method when multiple children need to send data to one parent in one cycle only. In data prediction, if a parent receives data, it again wakes-up after 3u, hoping that there will be data from another child as well. When same child needs to send multiple packets to same or different parent, a More-To-Send (MTS) flag is piggybacked in data-packet, so the parent will keep waking-up every after 3µ time, until it receives MTS flag set in last packet.

In-spite of using data-prediction and MTS flag, D-MAC is not suitable for high traffic load due to small µ time. D-MAC does not use RTS-CTS handshaking, because at a given time only few nodes of the network will remain active, thus reducing the chances of collision. Also, data aggregation is possible at each node because the parent can receive packets from all children before forwarding them. D-MAC requires local and efficient synchronization. Due to staggered wake-up schedule, each node should know its depth-level. We implemented D-MAC where the synchronization packet sinks down from root to leaf nodes informing each node about their depth level.  Another disadvantage is that D-MAC cannot be used to local-gossip type of communication due to its hierarchical data forwarding design.

## 5. B-MAC

Berkley Media Access Control (BMAC) [9] is a MAC level protocol for Wireless Sensor Networks which uses adaptive preamble sampling scheme. This technique consists of sampling the medium at fixed time intervals. Figure 4 describes the working of BMAC. Sampling the medium means to listen to the channel for some activity. In this scheme, every node samples the medium at fixed intervals to check whether any node is willing to communicate. If any node has a packet to send, node (sender) sense the medium if it is free, takes a small back-off and then sends a long wake up preamble followed by data packet. Preamble is not a packet but a physical layer RF pulse just greater than sampling period in length so that node sampling the medium notices this activity. However, for simplicity, we have considered the preamble as a long packet in our simulation. When receiver wakes up, it senses the medium and if it detects any noise (preamble), it turns on its radio and waits for the preamble to end. On completion of preamble, if data packet is destined to the node itself, it receives full data packet otherwise ignores the packet and goes to sleep.

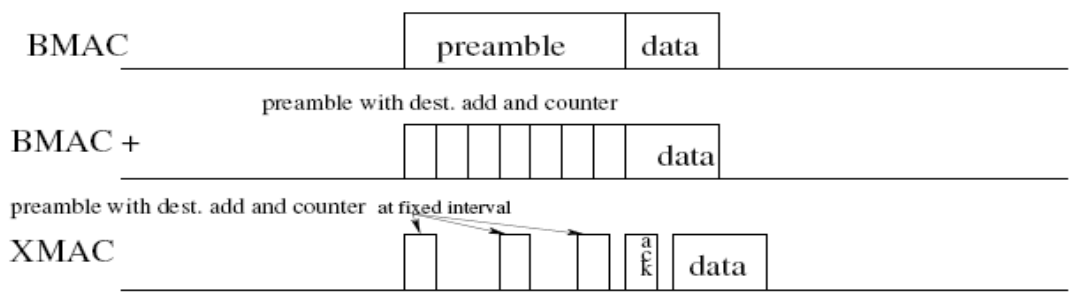

Figure 4. Preamble in BMAC, BMAC+ and XMAC

The goals of BMAC protocol are low power operation, effective collision avoidance and efficient channel utilization at low as well as high data rate. BMAC can be scaled to a large network. It is re-configurable by networks and its implementation is simple and requires small RAM size. BMAC protocol uses concepts of media access functionality. It uses clear channel assessment (CCA) and back offs for channel arbitration, acknowledgments for reliability, and low power listening (LPL) for low power communication. B-MAC is only a link protocol, with network services like organization, synchronization, and routing built above its implementation but BMAC is unable to provide multi-packet mechanisms like hidden terminal support, message fragmentation and particular low power policy.

## 6. B-MAC+

BMAC+ [10] is an extension of BMAC [9] protocol. BMAC+ tries to reduce waste of energy due to long preamble of BMAC. A Preamble is the sequence of bits which does not contain any relevant information and is used to tell receiver that some node wants to communicate. The basic idea of BMAC+ is to replace wake up preamble with small numbers of blocks containing some information as shown in Figure 4. This information contains address of destination node and number of remaining blocks or countdown of data starting with highest number according to preamble length needed. The destination address is used to avoid overhearing without receiving remaining preamble blocks. The number of remaining blocks or countdown of sequence number starting from zero is used to avoid idle listening by the nodes which are not recipients of data. BMAC+ saves more energy than BMAC with same latency and throughput as evident from simulation results in Figure 6. This is because in BMAC+, when receiver receives early preamble it can turn its radio off for remaining preamble blocks and wait for time of data arrival to turn radio on. It would again wake up when data arrives.

## 7. X-MAC

Standard MAC protocols such as BMAC use long preamble before data to wake up receiver. This preamble scheme was enhanced in BMAC+ to reduce power consumption. In BMAC receiver turns its radio off after getting preamble but still sender continue to send remaining part of preamble which results in waste of energy and also introduces excess latency.

In 2006, these problems were solved when XMAC [11], a low power MAC protocol was introduced. XMAC introduces a shortened preamble approach that retains the benefits of low power listening such as low power communication, simplicity and decoupling of transmitter-receiver sleep schedules. XMAC introduces a series of short preamble packets, each packet containing the destination address and remaining number of preambles. A node sends these series of preambles at fixed intervals as shown in Figure 4. This interval is long enough to get a response from receiver. When receiver receives any preamble, it at once sends acknowledgement to sender during fixed interval between preamble packets. When sender receives an acknowledgement, it stops sending further preambles and immediately sends data packet. This reduces both energy at both receiver and sender sides and also reduces latency per hop.

In addition, when a sender which is waiting for a clear channel to send data, detects a preamble and then hears an acknowledgement from a node to which sender itself wants to send data, it takes a random back-off which is long enough to complete data transmission currently going on. After completion of this transmission, it sends data directly without any preamble. The randomized back-off is necessary to avoid collision if more than one node is trying to send data at same time. Since this technique requires receiver node to remain awake to receive data in next cycle as well, every receiver in XMAC remains awake for a short period in case there are additional nodes willing to send data. These two techniques reduce a lot of energy consumption and also reduce latency.

## 7. W<small>ISE</small>-MAC

WiseMAC [12] is a medium access control protocol for WSNs which is based on non-persistent CSMA and uses preamble sampling technique to reduce power consumption. WiseMAC tries to use minimum sized wake preamble. WiseMAC requires no set-up signalling, no network-wide synchronization and is adaptive to traffic load. WiseMAC like BMAC is based on sampling technique where a node listens to the channel for a short duration. All sensor nodes sense medium at the same constant period 'TW' independently. Here 'independent' means that they may sense medium at different time but they sample the medium for same period. In BMAC, if the medium is found busy, the node continuously listens to the medium until data is received or until medium becomes idle again. At the sender side, wake-up preamble of a size equal to the sampling period is added in front of every data frame so that receiver will wake up at time of arrival of data packet. This protocol provides best result when medium is idle but disadvantage of this protocol is that long wake-up preambles cause a throughput limitation and large power consumption overhead in transmission and reception. The main idea of WiseMAC is learning the sampling schedules of direct neighbours of a node. These schedules are used to minimize size of preambles. To recover packet losses, a link level acknowledgement is used in WiseMAC. The WiseMAC ACK packets are not only used to carry acknowledgement information but also to inform other nodes (including sender) the remaining time of next sampling. These other nodes store this time in their tables. Using this information, a node transmits a packet with minimized size of preamble. The duration of the wake-up preamble covers the small potential clock drift between the clock at source and destination. WiseMAC uses following equality for calculating the minimum preamble. TP =min (4θL, TW), where θ is the frequency tolerance of time base quartz and L the interval between communications. L is also updated when a node gets ACK of its neighbours at Link level. The first communication between two nodes is always done using a long wake-up preamble equal to TW. Once some timing information is found, a wake up preamble of reduced size is used. Since preamble is proportional to the interval L between communications, it will become small when traffic is high. This makes WiseMAC adaptive to the traffic. The packet overhead decreases as per increase in traffic. In the low traffic conditions, the packet overhead is high, but average power consumption due to this overhead is low. Another important thing about WiseMAC is that it uses a more bit present in the header of data packet like IEEE 802.11 power save protocol. When this bit is set to 1, it indicates that more data is coming for same node. So, receiver continues checking the medium even after sending the acknowledgement. We have simulated WiseMAC in PowerTossim-Z and simulation results are shown in Figure 6. In figure we can see that energy consumption in WiseMAC is less than any other MAC protocol but greater than XMAC. Reason for this is that even though XMAC does not use minimized preamble but in XMAC all neighbouring nodes turn off their radio after receiving any single preamble while in WiseMAC, all neighbouring nodes also receive full preamble with actual receiver.

Thus sender's and receiver's energy may be saved but including all neighbouring nodes total energy is more than that of XMAC. Still WiseMAC performs better than others and may be more usable in low traffic conditions.

# 7. SIMULATION RESULTS

We evaluated all the above described MAC protocols over the parameters: delivery ratio, inter-arrival time, energy consumption and no. of hops. After implementation, protocols were integrated with PowerTOSSIM-Z [7] for measuring the energy consumption. Then each protocol was configured to give the delivery-ratio in the range 85-100%. For such a configuration energy consumption was evaluated. Note that, we did not integrate the MAC protocols with network layer, which uses Collection Tree Protocol (CTP) [13] due to conflict in beacon sending period of CTP and sleep periods of MACs. So a hard-coded hierarchical routing algorithm was manually written at the application layer for the evaluation purposes. A graph between average energy of each protocol and inter-arrival time is shown in Figure 5.

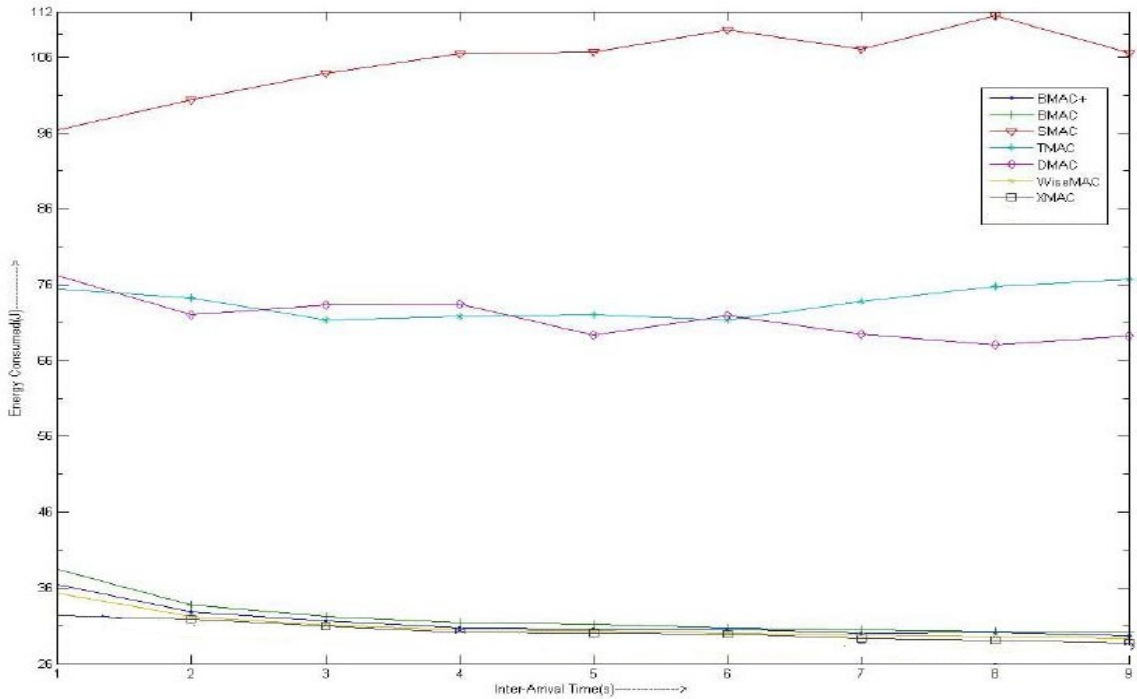

Figure 5. Energy consumption vs. Inter-arrival time for converge-cast communication

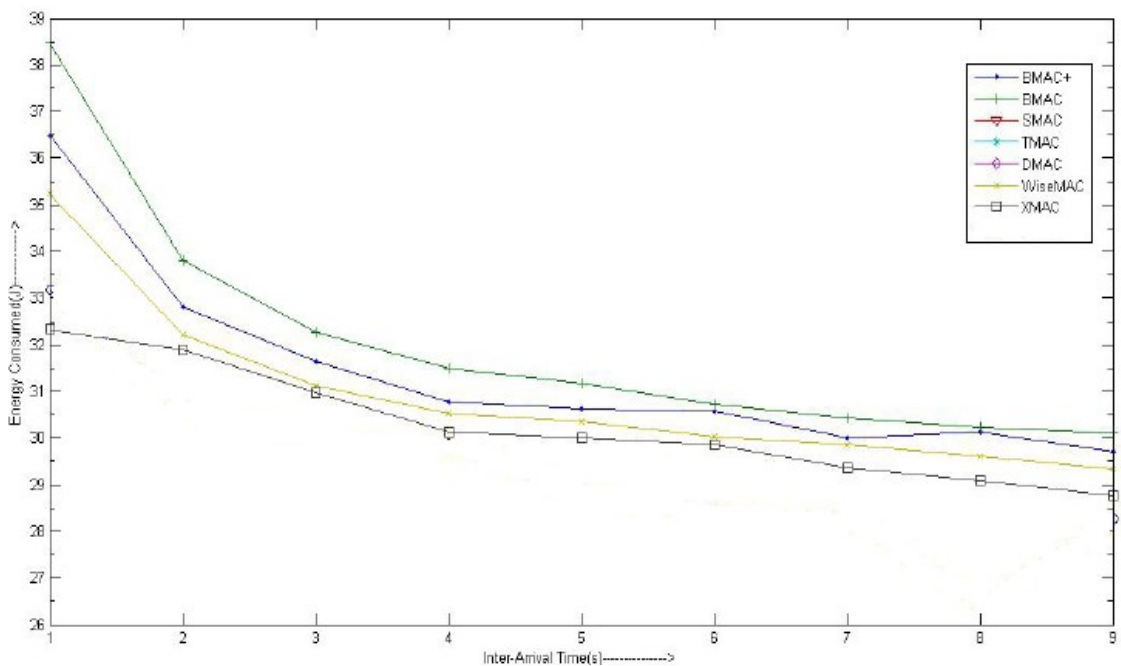

Figure 6. Closer view of analysis in Figure 5

A clean view of dense lines in figure 5 can be seen in figure 6.

In SMAC and TMAC, energy savings at high traffic is due to Overhearing avoidance. More neighbour nodes sleep for more time on hearing RTS and CTS when traffic is high, at low traffic they do not hear RTS/CTS thus wasting energy in idle listening during fixed duty cycle. Such observation is not seen in BMAC, XMAC, WiseMAC and DMAC. Thus, the order of energy consumption is:

XMAC<WiseMAC<BMAC+<BMAC<DMAC<TMAC<SMAC.

In this order if DMAC knows the levels of its node then DMAC can perform better than any MAC protocol. For knowing the level of nodes in DMAC, we need to pass a packet say SYNC packet to all nodes, which consume a lot of energy. In future, the work can be done on this part of DMAC.

Some of the protocols were tested for local gossip type of communication.

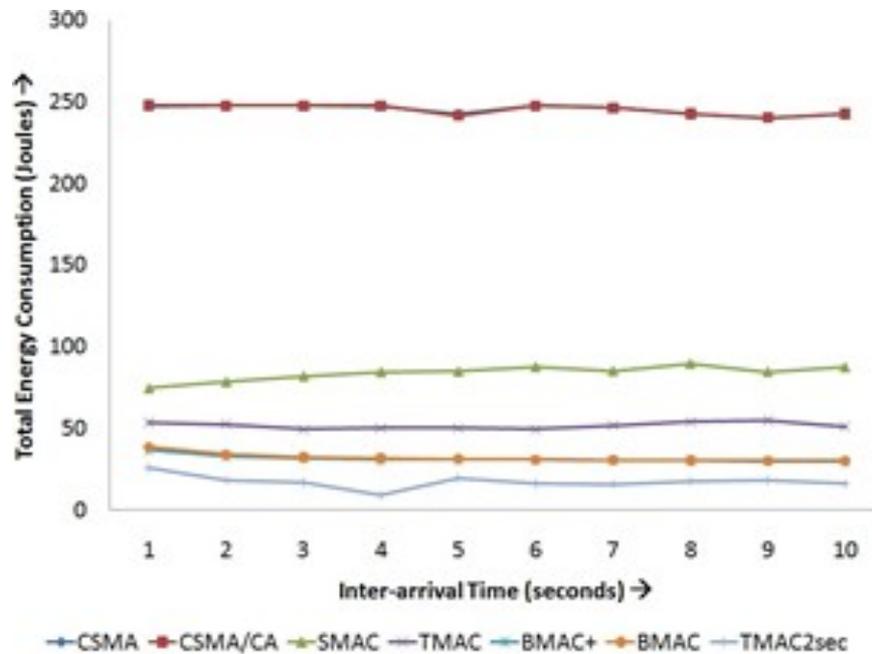

Figure 7. Energy Consumption for local gossip type of communication

Results shown in Figure 7 reveal that the energy consumption of TMAC was higher than that of SMAC for inter-arrival time 1 sec. This can attributed to the fact that TMAC can dynamically vary its active time according to data traffic. For high traffic rate, the active time (TA Time) may even become larger than that of SMAC. So energy consumption may be larger than SMAC for high data traffic, but decreases as data traffic decreases.

## 6. CONCLUSION

In this paper, we compared CSMA based MAC protocols with respect to their energy consumption and found that in general, X-MAC performs better than all other protocols which were considered. Protocols based on preamble sampling consume lesser energy than protocols based on static or dynamic sleep schedule. The paper also presented the advantages and disadvantages of these protocols when traffic is high and when it is low. Such analysis may be used to configure the network as per user requirements. In future, we aim to present a system with which users would be able to do such configurations using SQL like queries. We aim to integrate the implemented protocols with TinyDB2 [14], a query driven data extraction system for WSNs. The system would be integrated with PowerTossim-Z to enable users see the energy consumed by their application using SQL like queries without exposing them to internal details of sensor network platforms.

**Authors**

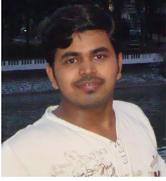
Himanshu Singh is a postgraduate student at Institute of Technology, B.H.U, India. He completed his Bachelors in Technology in Computer Engineering from Institute of Technology, B.H.U, India in 2011. His research interests include Computer Networks and Wireless Sensor Networks.

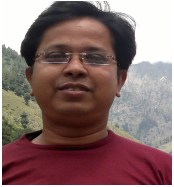
Bhaskar Biswas is an Assistant Professor at Department of Computer Engineering, Institute of Technology, Banaras Hindu University. He completed his PhD from IT-BHU, in 2011. His research interests include Database Systems, Data Mining and Web Mining.